\documentclass[apjl]{emulateapj}
\usepackage{apjfonts}

\newcommand{\mathsub}[2]{\ensuremath{\mathrm{#1_{#2}}}}
\newcommand{\cgsflux}{\ensuremath{\mathrm{erg \, cm^{-2} \, s^{-1}}}}

\slugcomment{Accepted to Ap.\ J.\ Letters 7 September 2005.}

\shorttitle{Knotty Question of the Jet of PKS B1421-490}
\shortauthors{Gelbord et al.}

\begin{document}



\title{The Knotty Question of the Jet of PKS B1421-490}

\author{J.~M. Gelbord\altaffilmark{1},
H.~L. Marshall\altaffilmark{1},
D.~M. Worrall\altaffilmark{2,3},
M. Birkinshaw\altaffilmark{2,3},
J.~E.~J. Lovell\altaffilmark{4},
R. Ojha\altaffilmark{4},
L. Godfrey\altaffilmark{4,5},
D.~A. Schwartz\altaffilmark{2},
E.~S. Perlman\altaffilmark{6},
M. Georganopoulos\altaffilmark{6,7},
D.~W. Murphy\altaffilmark{8},
D.~L. Jauncey\altaffilmark{4}}
\altaffiltext{1}{MIT Kavli Inst.\ for Astrophysics and Space
  Research, Massachusetts Inst.\ of
  Technology, 77 Massachusetts Ave., Cambridge, MA 02139, USA}
\altaffiltext{2}{Harvard-Smithsonian Center for Astrophysics,
  60 Garden St., Cambridge, MA 02138, USA}
\altaffiltext{3}{Dept.\ of Physics, U.\ of Bristol, Tyndall Ave.,
  Bristol BS8 1TL, UK}
\altaffiltext{4}{CSIRO ATNF,
  PO Box 76, Epping, NSW 2121, Australia}
\altaffiltext{5}{Research School of Astronomy \& Astrophysics,
 Australian National U.}
\altaffiltext{6}{Joint Center for Astrophysics and
  Physics Dept., U.\ of Maryland,
  Baltimore County, 1000 Hilltop Circle, Baltimore, MD 21250, USA}
\altaffiltext{7}{Goddard Space Flight Center, 
  Code 661, Greenbelt, MD 20771, USA}
\altaffiltext{8}{Jet Propulsion Lab., 4800 Oak Grove Dr.,
  Pasadena, CA 91109, USA}
\email{jonathan@\linebreak[0]space.mit.edu,
hermanm@\linebreak[0]space.mit.edu,
D.Worrall@\linebreak[0]bristol.ac.uk,
Mark.Birkinshaw@\linebreak[0]bristol.ac.uk,
Jim.Lovell@\linebreak[0]csiro.au,
Roopesh.Ojha@\linebreak[0]csiro.au,
lgodfrey@\linebreak[0]mso.anu.edu.au,
das@\linebreak[0]head-cfa.harvard.edu,
perlman@\linebreak[0]jca.umbc.edu,
markos@\linebreak[0]milkyway.gsfc.nasa.gov,
dwm@\linebreak[0]sgra.jpl.nasa.gov,
David.Jauncey@\linebreak[0]csiro.au
}

\begin{abstract}
We report the discovery of unusually strong optical and X-ray emission
associated with a knot in the radio jet of PKS~B1421-490.  
The knot is the brightest feature observed beyond the radio band,
with knot/core flux ratios $\sim$300 and 3.7 at optical and X-ray
frequencies.
We interpret the extreme optical output of the knot as synchrotron
emission.
The nature of the X-ray emission is unclear.
We consider a second synchrotron component, inverse Compton emission
from a relativistic, decelerating jet, and the possibility that this
feature is a chance superposition of an unusual BL~Lac object.
\end{abstract}


\section{Introduction}

We have observed PKS~B1421-490 \citep{Ekers69} as part of our
\textsl{Chandra} survey of flat-spectrum radio sources with extended
structure \citep{Marshall05}. 
A component at $\mathrm{14^h24^m32\fs23}$, --49\degr13\arcmin50\farcs0
(J2000.0) 
accounts for 93\%
of the 8.6~GHz flux density; the remaining emission extends
$\sim$12\arcsec\ to the south west \citep{Lovell97}.
\citet{Gelbord05b} discovered an optical counterpart with $g' = 24.2
\pm 0.2$ at the position of the radio peak.
We estimate the redshift to be $1 \la z \la 2$ by comparing the
$g'-r'$ and $r'-i'$ colors (0.46 and 0.30, respectively, after
dereddening) with a sample of SDSS quasars \citep{Richards02}.
An exceptionally bright optical feature ($g' = 17.8$) is
coincident with a weak component in the extended radio structure
5\farcs9 from the radio peak. 
Here we discuss this enigmatic feature in detail.

We adopt $z = 1$, $\mathsub{H}{o} = 70\ \mathrm{km \, s^{-1} \, Mpc^{-1}}$,
$\mathsub{\Omega}{m} = 0.3$, and $\mathsub{\Omega}{\Lambda} = 0.7$,
such that 1\arcsec\ corresponds to 8.0~kpc at the source.
We define the power law spectral index $\alpha$ by $S_\nu \propto
\nu^{-\alpha}$.
All uncertainties are 1-$\sigma$ and limits are 2-$\sigma$.

\section{Observations}

\begin{figure*}
\centering
%
\plotone{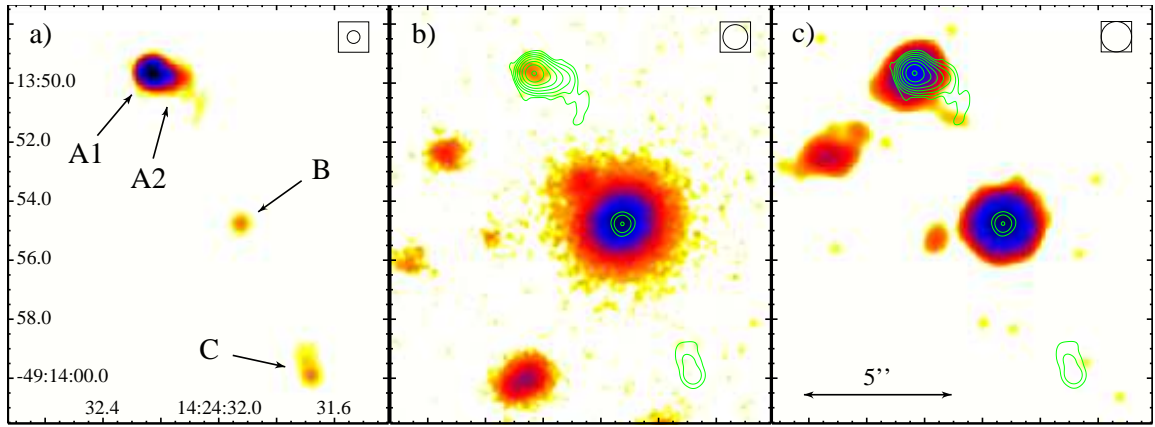}
\caption{PKS~B1421-490 imaged in radio, optical, and X-ray bands.
  a) \textsl{ATCA} radio map at 20.16~GHz, convolved to a circular beam
  with 0\farcs44 FWHM.  The map peaks at 2.14
  Jy/beam and has a background RMS level of 0.129~mJy/beam.
  b) \textsl{Magellan} $i'$ image (smoothed to 0\farcs86 FWHM)
  with radio contours corresponding to panel (a).
  The contour levels are separated by factors of three starting from
  $7 \times$ the background RMS.
  The astrometric solution, based upon 29 field stars 
  listed in the GSC 2.2.01,
  yields an offset of $<$0\farcs1 between the radio and optical
  centroids at A1.
  c) \textsl{Chandra} 0.5-7.0 keV image (convolved to 1\farcs03 FWHM)
  with overlaid radio contours.
  The logarithmic stretch spans 0.75--266 counts/beam.
  A shift by $\sim$0\farcs3 has been applied to register the
  northern peak with A1.
  All images are shown at the same angular scale, with the beam FWHM
  sizes indicated.
  Region B is slightly resolved in the radio map but 
  not in the optical or X-ray images, where it becomes the dominant
  component.
  The offsets between the radio, optical, and X-ray peaks at B are
  negligible ($<$0\farcs1).
  \label{allImages}}
\end{figure*}

We observed PKS~B1421-490 with the Australia Telescope Compact
Array (\textsl{ATCA})
at 4.86 and 8.64~GHz on 2002 February 4 and at 17.73 and 20.16~GHz on
2004 May 9.
The Magellan Instant Camera (MagIC) was used for imaging in the SDSS
$i'$, $r'$, and $g'$ filters on 2003 April 26--27,
and the Inamori Magellan Areal Camera and Spectrograph (IMACS)
was used to obtain a spectrum spanning 4100--7250~\AA\ 
during twilight on 2004 February 27.
X-ray data were obtained on 2004 January 16 using the \textsl{Chandra}
ACIS-S CCD. 
Imaging data are presented in
Fig.~\ref{allImages} and flux density ($S_\nu$) measurements 
in Tab.~\ref{fluxTab}.

%
%
%
\begin{deluxetable}{cccc}
\tablecaption{PKS~B1421-490 flux density measurements\label{fluxTab}}
\tablewidth{0pt}
\tablehead{
  \colhead{Frequency} &
  $S_{\nu, \mathrm{A}}$ \tablenotemark{a} & 
  $S_{\nu, \mathrm{B}}$ & 
  $S_{\nu, \mathrm{C}}$ \\
  \colhead{(Hz)} &
  (mJy) & 
  (mJy) & 
  (mJy)
}
\startdata
$4.86  \times 10^{9}$    & $4980	\pm 10$
                         & $< 7$
                         & $80		\pm 5$   		\\
$8.64  \times 10^{9}$    & $3679	\pm 7$
                         & $9.6		\pm 0.6$
                         & $46.1	\pm 0.6$ 		\\
$1.77  \times 10^{10}$   & $2100	\pm 4$
                         & $9.8		\pm 0.3$
                         & $20.6	\pm 0.3$		\\
$2.02  \times 10^{10}$   & $1960	\pm 4$
                         & $9.2		\pm 0.2$
                         & $17.5	\pm 0.3$		\\
$1.38 \times 10^{14}$    & $< 0.37$
                         & $1.00	\pm 0.07$
                         & $< 0.37$				\\
$1.82 \times 10^{14}$    & $< 0.27$
                         & $0.88	\pm 0.06$
                         & $< 0.31$				\\
$2.40 \times 10^{14}$    & $< 0.21$
                         & $0.91	\pm 0.07$
                         & $< 0.36$				\\
$3.93 \times 10^{14}$    & $(3.9	\pm 0.8) \times 10^{-3}$
                         & $0.81	\pm 0.12$
                         & $< 1.4 \times 10^{-3}$		\\
$4.82 \times 10^{14}$    & $(3.0	\pm 0.9) \times 10^{-3}$
                         & $0.78	\pm 0.15$
                         & $< 0.7 \times 10^{-3}$		\\
$6.29 \times 10^{14}$    & $(1.9	\pm 0.8) \times 10^{-3}$
                         & $0.73	\pm 0.20$
                         & $< 1.2 \times 10^{-3}$		\\
$2.41 \times 10^{17}$    & $(13.3	\pm 1.6) \times 10^{-6}$
                         & $(49 	\pm 3)	 \times 10^{-6}$
                         & \phn\phn $< 1.0 \times 10^{-6}$
\enddata
\tablecomments{Near-IR (from \textsl{2MASS};
  \citealp{2mass}) and optical 
  flux densities
  are corrected for Galactic extinction using
  A$_K=0.10$, A$_H=0.15$, A$_J=0.24$, 
  A$_{i'}=0.56$, A$_{r'}=0.73$, and A$_{g'}=1.01$
  \citep*{Schlegel98}.
  The optical flux uncertainties are dominated by
  this correction.
}
\tablenotetext{a}{At 18 and 20~GHz, region~A is resolved into A1 and
  A2.
  The tabulated flux densities refer to A1;
  $S_{\nu, \mathrm{A2}} = 600$ and 550~mJy at 18 and 20~GHz,
  respectively.
  At other frequencies the $S_{\nu, \mathrm{A}}$ values blend regions A1
  and A2.}
\end{deluxetable}
The radio structure exhibits three main emitting regions
(A, B, C on Fig.~\ref{allImages}a)
spanning 12\arcsec\ along position angle
$\mathrm{PA} \sim 209^\circ$ east of north.
The 18 and 20~GHz maps resolve region~A into a point source
and a jet-like 
extension (A1 and A2, respectively).
A2 has FWHM of $\sim$0\farcs56 by 0\farcs16, extending westward from
A1 ($\mathrm{PA} = 259$\degr) before bending
toward regions~B and C.
(Hereafter, unless A1 or A2 are specified, 
``region~A'' will refer to blends or the explicit sum of
these two subregions.)
At 2.3~GHz \citet{Preston89} report that A1 contains two VLBI
components with 
$\sim$30~mas diameters
separated by 55~mas 
roughly along the direction to A2.
An 8.425~GHz VLBI snapshot taken with the Long Baseline Array (LBA)
on 2004 April 16
%
%
shows a $\sim$20~mas jet with $\mathrm{PA} \sim 250$\degr,
but formally 
sets an upper limit of 24~mas on the size of this feature
within A1, with 
$\mathsub{S}{8.4\,GHz} = 1.3 \pm 0.3$~Jy.  
A1 has a flatter radio spectrum than A2.  Since it
includes some extended flux from the
base of A2 and certainly blends the VLBI structures,
there is likely to be a component within A1 with a flat or inverted
radio spectrum.

Region~B lies 5\farcs9 from A1 in $\mathrm{PA} = 211\degr$.
It is the weakest of the labeled radio features, with
only 0.4\% of the overall flux density of region~A at 20~GHz.
The 18 and 20~GHz data show it to be resolved 
(the deconvolved FWHM is $0\farcs12 \pm .03$ by
$0\farcs09^{+.03}_{-.05}$ with the major axis in $\mathrm{PA} =
21^{+21}_{-12}$ deg.). 
At 4.8~GHz C is sufficiently extended to 
contaminate 
region B.

Region~C is well resolved, reaching its
peak brightness at its southernmost
end, 11\farcs5 from A1 at $\mathrm{PA} = 208\degr$.
It has a steep spectrum ($\mathsub{\alpha}{r, C} = 1.15 \pm 0.03$
between 8.6 and 20~GHz),
in contrast to the flat spectra of A and B 
($\mathsub{\alpha}{r, A} = 0.448 \pm 0.005$ and $\mathsub{\alpha}{r,
  B} = 0.05 \pm 0.10$).
%

Above $10^{14}$~Hz (Fig.~\ref{allImages}b \& c), B
is the brightest component and C is undetected.
The \mathsub{S}{B}/\mathsub{S}{A} flux ratios are 3.7 in the X-ray
band and $\sim$300 in the optical, contrasting with $3 \times
10^{-3}$ at 8.6~GHz.
Both A and B are unresolved, with
FWHM upper limits of $0\farcs24$ at $10^{14}$~Hz,
and $0\farcs61$ and $0\farcs39$ (respectively) at $10^{17}$~Hz.
%
%
%
The spectral energy distribution (SED) of A appears to steepen in
the optical band ($\mathsub{\alpha}{o} = 1.49^{+0.70}_{-0.67}$),
while that of B remains flat ($\mathsub{\alpha}{o} = 0.22 \pm 0.23$).
Details of the optical data are given by \citet{Gelbord05b}.

The X-ray spectra of components A and B are both well described by
absorbed power laws.  
We use the maximum likelihood method to fit models to the
0.5--7.0~keV spectra.
For component B we find $\mathsub{\alpha}{x} =
0.42^{+0.24}_{-0.22}$ and 
a neutral gas column of 
$\mathsub{N}{H} = 1.0^{+0.7}_{-0.5} \times 10^{21}\ 
\mathrm{cm}^{-2}$,
consistent with the
\mathsub{N}{H} predicted from Galactic \ion{H}{1} measurements
\citep[$1.62 \times 10^{21}\ \mathrm{cm}^{-2}$;][]{Dickey90}.
For region~A we 
obtain $\mathsub{\alpha}{x} =
0.31^{+0.32}_{-0.31}$
after fixing $\mathsub{N}{H} = 1.62 \times 10^{21}\ \mathrm{cm}^{-2}$.

An unidentified optical and X-ray source (CXOU J142432.5-491352) lies
4\farcs1 SE of A1.
It has flux densities
of $9.2 \pm 1.5$, $7.5 \pm 1.7$, and $6.5 \pm 2.1~\mu$Jy in the $i'$,
$r'$, and $g'$ bands, respectively, and $3.1 \pm 0.8$~nJy at
1~keV.
The density of background X-ray sources at least this bright
($\mathsub{F}{2-10\,keV} = 3 \times 10^{-14}\ \cgsflux$,
assuming  
$\mathsub{\alpha}{x} = 0.6$ and Galactic \mathsub{N}{H})
is about 100 per square degree \citep{Rosati02,Moretti03}, so the
likelihood of
finding one within 10\arcsec\ of at least one of
our 30 survey targets 
observed to date
is $>$7\%.  The low Galactic latitude of
1421-490 ($b = 10.9\degr$)
increases the likelihood that this is an unrelated
source.
A weak optical source lies $1\farcs8$ NE of B ($g' = 23.1 \pm
0.3$, $r' = 22.5 \pm 0.2$, and $i' = 22.0 \pm 0.2$ with no extinction
correction).
It lacks any radio or
X-ray counterpart and is consistent with a point source,
hence is likely to be a foreground star in this crowded field.
We do not discuss either object further.

\section{Deciphering the morphology}

\label{sectMorph}
We suggest that region C is a terminal hot spot due to its extended
radio structure and steep spectrum.
The knotty question is whether we have 
a core at A with a one-sided jet extending through B to C 
or a symmetric system with a core at B between 
hot spots at A and C.
The radio spectral indices do not provide guidance
because both A and B have flat spectra typical of
self-absorbed quasar cores.
The detection of a compact, high brightness temperature VLBI source 
with apparent core-jet morphology 
coincident with A1 
is suggestive but not conclusive:
a hot spot could contain a compact component,
although it would be exceptional for
it to have core-jet structure and
to provide so much (35\%) of the $\sim$8.5~GHz flux density.
The lack of a VLBI source at B is likewise not
significant because B would not have been detected by
extant observations.
For now we adopt the one-sided jet scenario, guided by the VLBI source
at A1.
This choice may need to be reconsidered as new data become available.
However, we note that 
$\mathsub{\alpha}{r, A1} < \mathsub{\alpha}{r, A2}$,
consistent with a core/inner jet interpretation.

One other possibility warrants consideration: 
that A and B are cores in separate systems.
In comparison with other jet knot-core
pairs, the B/A X-ray flux ratio is atypical and the optical
ratio unprecedented.
Given that X-ray sources with
$\mathsub{F}{2-10\,keV} \ge 4.3 \times 10^{-13}\ \cgsflux$ have a
density of $\sim$1.0
per square degree \citep{Moretti03}, 
%
%
the chance of finding such a source $<$10\arcsec\ from any  
of the  
systems so far observed in our jet survey 
is $\sim$$7.3 \times 10^{-4}$ 
if we neglect the
possibility of clustering.
Thus it is unlikely that A and B are unrelated.

%
\begin{figure}
\centering
\plotone{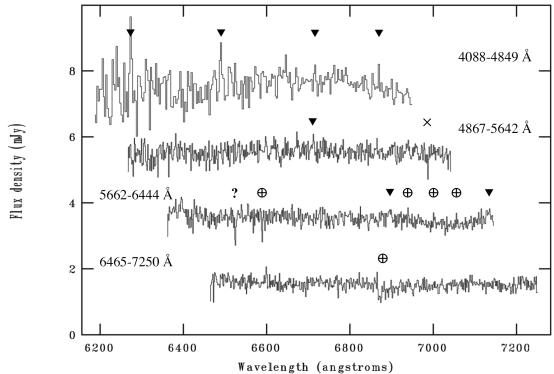}
\caption{IMACS spectrum of region B.
  From the top down the spectral segments have been shifted by 2100,
  1400, 700 \& 0~\AA\ and by 6, 4,  2 \& 0~mJy.
  Residual telluric features ($\oplus$), cosmic rays
  ($\blacktriangledown$), and a bad column at 
  5587 \AA\ ($\times$) are marked.
  The spectral resolution is $\Delta \lambda = 3.5$~\AA\ except for
  the bluest segment which is rebinned to $\Delta \lambda = 7.0$~\AA.
  The spectrum has been dereddened assuming A$_V =
  0.884$ \citep{Schlegel98}.
  The most
  significant feature is a tentative absorption line at 5825~\AA\ 
  (marked with ``?''),
  which represents a dip from the continuum by $4.3 \times$ the local
  RMS.
  A Lyman forest can be ruled out for $\lambda > 4370$~\AA ,
  confirming that $z < 2.6$.
  \label{optSpec}} 
\end{figure}
We obtained an optical spectrum (Fig.~\ref{optSpec})
to test whether B is an interloper.
Only one tentative feature, an absorption line at 5825\AA, exceeds $4
\times$ the local RMS.
A redshift limit of $z < 2.6$ is required by the absence of a Lyman
forest at $\lambda > 4370$\AA.
The lack of strong spectral features rules out identification with
common objects such as normal stars, galaxies and most types of AGN, 
and is consistent with non-thermal emission from a jet knot.
%
The only remaining candidates are DC white dwarfs (WDs)
and BL~Lac objects.
The former can be dismissed immediately as these
cool WDs        
are neither radio nor X-ray
sources and are generally bluer
($r'-i' < 0.00$, \citealp[e.g.,][]{Kleinman04},
whereas B has $r'-i' = 0.21$).
%
%
Moreover, 
photographic plates from the Yale/San Juan Southern Proper Motion
survey \citep[SPM;][]{Girard04}
limit the proper motion of B to $<$8 mas yr$^{-1}$ 
(D.~I.\ Dinescu 2004, private communication); with a distance
$<$150~pc (based upon the brightest known DC WDs; \citealp{McCook99})
the tangential motion is limited to 
$T < 6 \ \mathrm{km \, s^{-1}}$ 
whereas WDs typically have $T = 55 \pm 45 \ \mathrm{km \, s^{-1}}$
\citep{Sion88}.

If B is a BL~Lac, then it is highly unusual.
Its \mathsub{\alpha}{rx} index is typical of X-ray-selected
BL~Lacs, but \mathsub{\alpha}{ro} and
\mathsub{\alpha}{ox} ($< 0.20$ and $1.62$,
respectively\footnote{Evaluated with the upper limit
  at 4.8~GHz and extrapolated $S_\nu$ values at 2500~\AA\ and
  2~keV to facilitate comparison with \citealp{Rector00}.})
are 
exceptional, making B an outlier when compared to published samples
\citep{Worrall99,Rector00,Landt01}.
However, other methods of selection may broaden the 
distribution of indices--- 
e.g., the optically-selected,
radio-quiet BL~Lac candidate discovered by \citet{Londish04}.
BL~Lac objects typically vary in the optical by a
magnitude or more on time scales of years, but 
catalogs and archival data\footnote{The USNO-A2.0 \citep{Usnoa2},
  USNO-B1.0 \citep{Usnob1} and HST GSC 2.2
  (http://\linebreak[0]www-gsss.stsci.edu/\linebreak[0]gsc/\linebreak[0]GSChome.htm)
  catalogs, and the SPM survey
  (D.~I.\ Dinescu 2004, private communication).}
show $\Delta \mathsub{B}{j} < 0.6$ over
35~yr and $< 0.3$ on time scales up to 27~yr.
Finally, BL~Lacs with comparable X-ray fluxes are rare:
the chance of having one
within 10\arcsec\ of {\em any} of our 30 \textsl{Chandra}
targets is $< 1.5 \times 10^{-4}$ \citep{Wolter91,Krautter99,Henry01};
%
%
finding one aligned with a radio jet is even less likely.
%
%
Thus, while we cannot completely rule out a BL~Lac interloper, we
favor an association of B with PKS~B1421-490. 

The multiwavelength properties of feature B are unlike those of any
other known jet knot.
Its optical apparent magnitude is second only to knot HST-1 of M87
despite lying at a much greater distance.
HST-1 has a comparable X-ray \mathsub{S}{knot}/\mathsub{S}{core}
ratio,
but its optical ratio is 3, compared to $\sim$300 for knot~B
\citep{Marshall02,Harris03}.

\section{Interpreting the spectra}

%
\begin{figure}
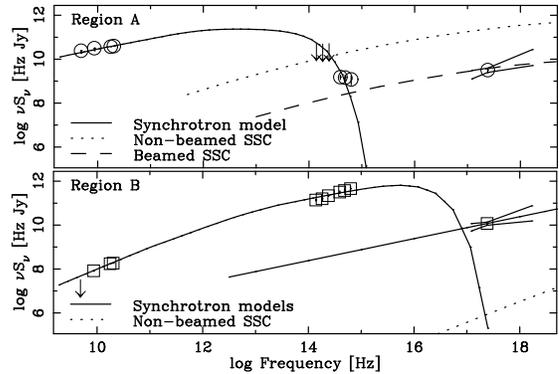

\centering
\plotone{figure3a.eps}
\plotone{figure3b.eps}
\caption{
  Spectral energy distributions (SED) and models of the core
  (region A; {\em top}) and knot B ({\em bottom}).
  Synchrotron models ({\em solid curves}) are drawn through
  the radio and optical fluxes of both sources.
  For the core, a synchrotron self-Compton (SSC) model with
  significant bulk relativistic motion ({\em dashed line}) gives a
  good fit to the X-ray flux,
  whereas a non-relativistic SSC model ({\em dotted line}) is
  too high.
  For the knot, the predicted SSC contribution is far too low; 
  it would be suppressed further if a 
  high \mathsub{\gamma}{min} or a hardened electron distribution
  is invoked to accommodate the \mathsub{S}{4.8 \mathrm{GHz}} limit.
  An \textit{ad hoc} second synchrotron component with a typical slope
  ($\alpha = 0.5$) could work, matching the X-ray data while
  contributing negligibly at lower frequencies.
  \label{sedFig}}
\end{figure}
The core SED can be modeled as emission from the parsec-scale base of
the jet.
The spectral index of A1 between 17.7 and 20.2~GHz is 0.5, roughly
consistent with strong-shock acceleration models.
A synchrotron self-Compton (SSC) model with an equipartition
magnetic field ($B_{\rm eq}$) is plausible for the core as long as there is
relativistic boosting towards the observer.
For example, Fig.~\ref{sedFig} 
shows a model of a jet with bulk motion of $\Gamma = 20$ at 2.9\degr\
to the line 
of sight, radius 0.5~mas (4~pc), electron number spectrum 
index $p = 2.0$,
$\mathsub{\gamma}{min} = 20$, $\mathsub{\gamma}{max} = 1.3 \times
10^4$, and an energy-loss break by 1.0 
at $\gamma = 1.6 \times 10^3$.
Such a model gives a reasonable
description of the SED with 
$B_{\rm eq} = 13$~mG. 
Without relativistic beaming SSC models with $B \approx
B_{\rm eq}$ predict an X-ray flux $\sim$$70 \times$ higher.

For region B we also use synchrotron emission to model the
radio-to-optical flux.
The flat spectrum and apparent turn-over at low radio frequencies
might suggest self-absorption in a series of small-scale components.  
This is often the case in cores but not large-scale jets.
While there could be a self-absorbed region within the kpc-scale
knot if the magnetic field is sufficiently high ($\sim$mG),
we prefer instead an optically thin model.
Without beaming, a model consisting of a sphere of radius
0\farcs05 (0.4~kpc), with $B_{\rm eq} = 0.86$~mG, 
$p = 1.0$, 
$\mathsub{\gamma}{min} = 20$, 
$\mathsub{\gamma}{break} = 1 \times 10^5$, 
and $\mathsub{\gamma}{max} = 3 \times 10^6$,
yields a reasonable fit (Fig.~\ref{sedFig}),
although it violates the \mathsub{S}{4.8 \mathrm{GHz}} limit and
it is difficult to reproduce
$\mathsub{\alpha}{o} < 0.5$ 
unless $p < 2.0$ above \mathsub{\gamma}{break}.
An unusually high
\mathsub{\gamma}{min} ($\sim 2 \times 10^4$) would provide a better
compromise between the flat 8.6--20~GHz radio spectrum and the low
4.8~GHz flux density,
although the physical mechanism would be uncertain;
$B_{\rm eq}$ would be little changed.
%
The flatness of the optical spectrum suggests that the electron
distribution 
would be better modelled as multiple components with
a distribution of \mathsub{\gamma}{max} values.
The radiative lifetime
of the optically-emitting electrons
is only of order 40~years, implying distributed
electron acceleration across the knot.
Smaller electron Lorentz factors apply if the source size
is reduced.

The greater difficulty is to explain the X-ray emission.
Unlike the core, the relatively larger angular size and lower radio
luminosity of the knot means that SSC is not effective unless most of
the radio emission is from embedded regions smaller than 
we adopt for the core.
The observed flux is too low to be a simple continuation
of the optical synchrotron and too flat to be the plunging
high-energy tail of this component.
Synchrotron X-ray emission requires
a second population of electrons
extending to $\gamma \gtrsim 10^7$
with $1.4 \lesssim p \lesssim 2.3$.
The resulting emission
extrapolates well below the fluxes
measured at lower frequencies,
thus \mathsub{\gamma}{min} is unconstrained.  
{\em In situ} acceleration is required to
sustain this population as the cooling time
is much shorter than the light-crossing time of the region.

Inverse Compton (IC) models likewise have problems reproducing the
knot X-ray emission.
IC scattering of the cosmic microwave background
(IC-CMB; \citealp{Tavecchio00}; \citealp*{Celotti01})
for $B = B_{\rm eq}$ requires a bulk Lorentz factor $\Gamma > 60$ to
boost the X-ray output by 8 orders of magnitude.
This is inconsistent with our expectation 
that the knot should be less boosted than the core, 
especially after a bend in the jet as between A2 and B.
A higher assumed redshift would ease this constraint, since
$\Gamma \propto (1+z)^{-(3+\alpha)/(1+\alpha)}$ \citep{Harris02}, 
but even at $z = z_{\rm max} = 2.6$ we require $\Gamma > 24$.
This limit could be reduced to $\Gamma \lesssim 20$ if
magnetic fields significantly below $B_{\rm eq}$ are considered.

A decelerating jet model
\citep{Georganopoulos03a} 
with a lower $\Gamma$ and $B \approx B_{\rm eq}$ 
can provide a qualitatively correct SED.
A $\Gamma = 20$, $B = 0.1$~mG flow oriented 2.9\degr\ from our line of
sight that decelerates to semi-relativistic velocities within the $R =
0.4$~kpc volume of knot B reproduces the optical flux with a strongly
Doppler-boosted synchrotron peak and provides X-rays through IC
scattering of photons from the slowed, downstream portion of the jet.

Both IC models require abundant $\gamma \sim 100$ electrons
to produce X-rays.
These electrons can be in a second population (this time at low
energies), or part of the synchrotron-emitting population if
\mathsub{\gamma}{min} is much lower than the assumed value of $1.6
\times 10^4$.
A low \mathsub{\gamma}{min} would require the knot to contain
unusually compact, self-absorbed sub-regions to avoid producing an
order of magnitude more radio flux than observed.

\ \ 

\ \

\section{Conclusions}

We have chosen to interpret PKS~B1421-490 as a flat spectrum radio
source with a one-sided jet and a unique knot at B.
The optically-dominated spectral energy distribution of feature~B 
can be explained by two models.
Both invoke a synchrotron component for the radio-to-optical
continuum, but one attributes the X-ray flux to synchrotron
emission from a second electron population while the other involves
inverse Compton scattering by a relativistic, decelerating jet.
Alternative interpretations that currently cannot be ruled out include
a symmetric system with a core at~B and knots at~A and~C, and either an
interaction, or a chance alignment, of PKS~B1421-490 and an
unrelated (and unusual) optically-dominated, radio-quiet BL~Lac.

Deeper optical spectroscopy is urgently needed to measure the redshift
of A and provide a high S/N spectrum to better identify B.
Observations in the mm, sub-mm, IR and far-UV bands would fill in the
SED, thereby providing strong constraints for models of the emission
processes.
Feature B is 
a good target for optical
polarimetry, which could help confirm whether it is 
dominated by synchrotron jet emission.
Finally, we look forward to an upcoming long \textsl{Chandra}
observation that will allow a more detailed X-ray study 
and a deep VLBI mapping that will test the core-jet structure at A1
and possibly identify compact emission regions within B.

\ \ 

\acknowledgements

The authors thank Bill van Altena, Dana Dinescu and Terry Girard for
their assistance with the SPM archive.
This work has been supported in part 
under SAO contracts GO4-5124, SV1-61010, and NAS8-39073, and NASA LTSA
grant NAG5-9997.
The ATCA and the LBA are part of the Australia Telescope which is
funded by the Commonwealth of Australia for operation as a National
Facility managed by CSIRO.
This research has made use of 
the NASA/IPAC Extragalactic Database (NED),
NASA's Astrophysics Data System (ADS),
and data from the HST Guide Star Catalog
(GSC,
produced at
STScI
under U.S.\ Government grant).


\begin{thebibliography}{30}
\expandafter\ifx\csname natexlab\endcsname\relax\def\natexlab#1{#1}\fi

\bibitem[{{Celotti} {et~al.}(2001){Celotti}, {Ghisellini}, \&
  {Chiaberge}}]{Celotti01}
{Celotti}, A., {Ghisellini}, G., \& {Chiaberge}, M. 2001, \mnras, 321, L1

\bibitem[{{Cutri} {et~al.}(2003)}]{2mass}
{Cutri}, R.~M., {et~al.} 2003, VizieR Online Data Catalog,
  2246, 0, {(2MASS All-Sky Catalog)}

\bibitem[{{Dickey} \& {Lockman}(1990)}]{Dickey90}
{Dickey}, J.~M., \& {Lockman}, F.~J. 1990, \araa, 28, 215

\bibitem[{{Ekers}(1969)}]{Ekers69}
{Ekers}, J.~A. 1969, {Au.\ J.\ of Phys.\ Astrophys.\ Supp.}, 7, 3

\bibitem[{{Gelbord} \& {Marshall}(2005)}]{Gelbord05b}
{Gelbord}, J.~M., \& {Marshall}, H.~L. 2005, in prep.

\bibitem[{{Georganopoulos} \& {Kazanas}(2003)}]{Georganopoulos03a}
{Georganopoulos}, M., \& {Kazanas}, D. 2003, \apjl, 589, L5

\bibitem[{{Girard} {et~al.}(2004)}]{Girard04}
{Girard}, T.~M., {Dinescu}, D.~I., {van Altena}, W.~F., {Platais}, I., {Monet},
  D.~G., \& {L{\' o}pez}, C.~E. 2004, \aj, 127, 3060

\bibitem[{{Harris} {et~al.}(2003)}]{Harris03}
{Harris}, D.~E., {Biretta}, J.~A., {Junor}, W., {Perlman}, E.~S., {Sparks},
  W.~B., \& {Wilson}, A.~S. 2003, \apjl, 586, L41

\bibitem[{{Harris} \& {Krawczynski}(2002)}]{Harris02}
{Harris}, D.~E., \& {Krawczynski}, H. 2002, \apj, 565, 244

\bibitem[{{Henry} {et~al.}(2001)}]{Henry01}
{Henry}, J.~P., {Gioia}, I.~M., {Mullis}, C.~R., {Voges}, W., {Briel}, U.~G.,
  {B{\" o}hringer}, H., \& {Huchra}, J.~P. 2001, \apjl, 553, L109

\bibitem[{{Kleinman} {et~al.}(2004)}]{Kleinman04}
{Kleinman}, S.~J., {et~al.} 2004, \apj, 607, 426

\bibitem[{{Krautter} {et~al.}(1999)}]{Krautter99}
{Krautter}, J., {et~al.} 1999, \aap, 350, 743

\bibitem[{{Landt} {et~al.}(2001)}]{Landt01}
{Landt}, H., {Padovani}, P., {Perlman}, E.~S., {Giommi}, P., {Bignall}, H., \&
  {Tzioumis}, A. 2001, \mnras, 323, 757

\bibitem[{{Londish} {et~al.}(2004)}]{Londish04}
{Londish}, D., {Heidt}, J., {Boyle}, B.~J., {Croom}, S.~M., \&
  {Kedziora-Chudczer}, L. 2004, \mnras, 352, 903

\bibitem[{Lovell(1997)}]{Lovell97}
Lovell, J.~E.~J. 1997, PhD thesis, University of Tasmania

\bibitem[{{Marshall} {et~al.}(2002)}]{Marshall02}
{Marshall}, H.~L., {Miller}, B.~P., {Davis}, D.~S., {Perlman}, E.~S., {Wise},
  M., {Canizares}, C.~R., \& {Harris}, D.~E. 2002, \apj, 564, 683

\bibitem[{{Marshall} {et~al.}(2005)}]{Marshall05}
{Marshall}, H.~L., {et~al.} 2005, \apjs, 156, 13

\bibitem[{{McCook} \& {Sion}(1999)}]{McCook99}
{McCook}, G.~P., \& {Sion}, E.~M. 1999, \apjs, 121, 1

\bibitem[{{Monet} {et~al.}(1998)}]{Usnoa2}
{Monet}, D.~B.~A., {et~al.} 1998, {The
  USNO-A2.0 Catalogue} (Washington, DC: {U.S. Naval Observatory})

\bibitem[{{Monet} {et~al.}(2003)}]{Usnob1}
{Monet}, D.~G., {et~al.} 2003, \aj, 125, 984

\bibitem[{{Moretti} {et~al.}(2003)}]{Moretti03}
{Moretti}, A., {Campana}, S., {Lazzati}, D., \& {Tagliaferri}, G. 2003, \apj,
  588, 696

\bibitem[{{Preston} {et~al.}(1989)}]{Preston89}
{Preston}, R.~A., {et~al.} 1989, \aj, 98, 1

\bibitem[{{Rector} {et~al.}(2000)}]{Rector00}
{Rector}, T.~A., {Stocke}, J.~T., {Perlman}, E.~S., {Morris}, S.~L., \&
  {Gioia}, I.~M. 2000, \aj, 120, 1626

\bibitem[{{Richards} {et~al.}(2002)}]{Richards02}
{Richards}, G.~T., {et~al.} 2002, \aj, 123, 2945

\bibitem[{{Rosati} {et~al.}(2002)}]{Rosati02}
{Rosati}, P., {et~al.} 2002, \apj, 566, 667

\bibitem[{{Schlegel} {et~al.}(1998){Schlegel}, {Finkbeiner}, \&
  {Davis}}]{Schlegel98}
{Schlegel}, D.~J., {Finkbeiner}, D.~P., \& {Davis}, M. 1998, \apj, 500, 525

\bibitem[{{Sion} {et~al.}(1988)}]{Sion88}
{Sion}, E.~M., {Fritz}, M.~L., {McMullin}, J.~P., \& {Lallo}, M.~D. 1988, 
  \aj, 96, 251

\bibitem[{{Tavecchio} {et~al.}(2000)}]{Tavecchio00}
{Tavecchio}, F., {Maraschi}, L., {Sambruna}, R.~M., \& {Urry}, C.~M. 2000,
  \apjl, 544, L23

\bibitem[{{Wolter} {et~al.}(1991)}]{Wolter91}
{Wolter}, A., {Gioia}, I.~M., {Maccacaro}, T., {Morris}, S.~L., \& {Stocke},
  J.~T. 1991, \apj, 369, 314

\bibitem[{{Worrall} {et~al.}(1999)}]{Worrall99}
{Worrall}, D.~M., {Birkinshaw}, M., {Remillard}, R.~A., {Prestwich}, A.,
  {Tucker}, W.~H., \& {Tananbaum}, H. 1999, \apj, 516, 163

\end{thebibliography}
\end{document}